# Detections of He-3 in Ni-based binary metal nanocomposites with Cu in zirconia exposed to hydrogen gas at elevated temperatures

Tomoya Yamauchi [1], Yutaka Mori [1], Shuto Higashi [1], Hayato Seiichi [1], Masahiko Hasegawa [1], Akito Takahashi [2], Akira Taniike [1], Masato Kanasaki [1*]

[1] Graduate School of Maritime Sciences, Kobe University, Kobe, 658-0022, Japan
[2] Osaka University, Suita, Osaka 565-0871, Japan
E-mail: kanasaki@maritime.kobe-u.ac.jp

Abstract

The present study aims to detect helium-3 in nickel-based metal nano-composites doped with zirconia, which exhibited anomalous heat generation when exposed to hydrogen gas at approximately 450°C. Two complementary analytical techniques were employed: Nuclear Reaction Analysis (NRA) utilizing 1.4 MeV deuteron beams from a tandem accelerator, and Thermal Desorption Spectrometry (TDS) using a quadrupole mass spectrometer. Both methods successfully detected helium-3 in the samples. Given the extreme rarity of this isotope, its presence strongly suggests the occurrence of nuclear reactions within the nickel-containing materials. These findings lend support to the 4H/TSC (4 Hydrogen/Tetrahedral Symmetric Condensate) model, which uniquely predicts helium-3 as one of the primary reaction products.

## 1. Introduction

This investigation represents the first endeavor to substantiate the occurrence of nuclear reactions by detecting helium-3 in nickel-based binary metal nanocomposites with copper doped in zirconia, following exposure to hydrogen gas at elevated temperatures. The research falls within the domain of "condensed matter nuclear science" or "cold fusion," a field inaugurated in 1989 by Fleischmann and Pons' groundbreaking work on heavy water electrolysis (for a comprehensive review, see [1]). Recent advancements in this field have been propelled by the development of nanostructured multilayer metal composites and hydrogen gas interactions at Tohoku University, as reported by Iwamura *et al.*, (2024).

Atmospheric helium exhibits a He-3 abundance of approximately one part per million [2], a composition attributed to primordial nucleosynthesis during the cosmic Big Bang [3]. The extreme scarcity of He-3 implies that its detection in mere hundreds of milligrams of material would constitute compelling evidence for the occurrence of nuclear reactions, potentially including light hydrogen "cluster"-fusion processes.

One theoretical model that predicts the generation of He-3 is the 4H/TSC (4 Hydrogen/Tetrahedral Symmetric Condensate) WS (weak and strong nuclear cascade fusion) model [4, 5]. The concepts of the model can be summarized as follows: if we consider the problem of symmetrically placing four protons and four electrons at the eight vertices of a cube, so that the electrons and protons are located at the shortest distance corresponding to the twelve edges of the cube, we obtain a tetrahedron created by the four protons and the accompanying and orthogonally locating tetrahedron by the four electrons. The electrostatic potentials of these protons and electrons as classical point charges lead to the result that the smaller the cube, the more stable this system is. Takahashi [4,5] has solved the dynamic condensation process of the system of protons and electrons using quantized Langevin equations. As a result of the weak interaction (electron capture) and the immediate strong interaction, the system will become an unstable Li*-4 nucleus. Two cases are predicted for the decay of Li*-4: one producing He-3 and a proton ($^4$Li* → $^3$He + p + 7.73 MeV), the other producing a deuteron and two protons ($^4$Li* → d + 2p + 2.22 MeV). The branching ratio of the two decay processes will be experimentally determined. We chose to focus on the He-3 as a consequence, rather than continuing the debate on whether or not tetrahedral structures could occur as initial conditions in or on the surface of nano-sized Ni crystal lattices. Provided that He-3 were detected, this would provide evidence in favor of the 4H/TSC model. Of course, this does not rule out the future possibility of other models describing He-3 production.

While previous studies have sought to capture helium as a product of cold fusion reactions [6], this investigation represents a novel approach in specifically targeting He-3. The research's innovative premise stems from the hypothesis that He-3 is entrapped within the Cu-Ni/zirconia compound matrix. To explore this hypothesis, two distinct analytical techniques were employed: Nuclear Reaction Analysis (NRA) utilizing deuteron beams from an accelerator, and Thermal Desorption Spectrometry (TDS) using a quadrupole mass spectrometer coupled to a vacuum chamber with a controllable heating element. The Cu-Ni/zirconia compound materials were integral to the feasibility of this research endeavor.

## 2. Materials

The materials analyzed in this study were nickel-based binary metal nanocomposites with minor copper content, supported in a zirconia matrix. These samples were exposed to hydrogen gas at pressures of approximately 300 kPa and temperatures around 450°C. The Cu-Ni/zirconia nanocomposite was developed during a NEDO (New Energy and Industrial Technology Development Organization) project from 2015 to 2017, involving a collaborative effort among research groups from Tohoku University, Nagoya University, Kyushu University, Nissan Motor Co., Technova Inc., and Kobe University [7].

The fabrication process involved three key steps: First, thin amorphous metal ribbons of $Cu_xNi_yZr_z$ composite alloys, 10 μm in thickness, were produced via melt-spinning. Second, these ribbons underwent calcination in air at 450°C for 60 hours, resulting in the preferential formation of a $ZrO_2$ matrix with isolated $Cu_xNi_y$ nanostructured zones. Finally, the oxidized ribbons were ground to particles ranging from several to tens of micrometers in diameter.

The NEDO project concluded with the confirmation that several materials, including this Cu-Ni/zirconia composite, exhibited anomalous heat generation in hydrogen gas at elevated temperatures, which could not be attributed to known chemical reactions [7]. Subsequent research by Technova Inc. and Kobe University has led to significant improvements in the exothermic properties of these materials [8-12].

A crucial technique for enhancing exothermic properties was the implementation of re-calcination—a heat treatment process in which the Cu-Ni/zirconia sample is maintained in air at 450°C for a specified duration following exothermic experiments in hydrogen gas. Repeated re-calcination was observed to further improve exothermic properties, though detailed characterization of recent advancements in heat output will be reported separately. Table 1 provides a comprehensive overview of the samples used in the present analysis, including their initial composition, heat treatment and exothermic experiment history, integrated calorific value per 0.1 g, and the analytical methods employed.

Table 1 List of Cu-Ni/zirconia samples analyzed.

| Sample ID | Composition ratio Cu:Ni:Zr | History of heat treatments and experiments | Integrated calorific value (J/0.1 g) | Analysis method |
|---|---|---|---|---|
| #1 | 1:10:20 | A+B+C+B+C+B | $7.29 \times 10^3$ | NRA |
| #2 | 1:10:20 | A+B+C+B+C+B'+C | $3.20 \times 10^4$ | NRA |
| #3 | 1:7:14 | A+B+C+B+C+B+C | $5.50 \times 10^4$ | NRA |
| #4 | 1:7:14 | A+B+C+C+C | $1.39 \times 10^4$ | TDS |

A: Calcination at 450°C for 60 h and heat generation experiments in $H_2$ gas.

B: Re-calcination at 450°C for 180 h    B': Re-calcination at 450°C for 60 h

C: Heat generation experiments in $H_2$ gas.

## 3. Preliminary considerations and experiments

### 3.1. Validity of the present hypothesis

The samples of Cu-Ni/zirconia should be regarded as consisting of two phases of Ni-Cu alloy and $ZrO_2$ matrix. A review paper on the behavior of helium in inorganic and metallic materials, compiled from the perspective of nuclear materials research, is reported by Trocellier et al. [13]. Helium retention properties have been investigated for Cu and Ni and $ZrO_2$ respectively. The methods we employed to test our 'hypothesis', which are also introduced in this review paper, are Thermal Desorption Spectrometry, TDS, and Nuclear Reaction Analysis, NRA. In TDS, the sample is heated in a vacuum at a constant heating rate and the partial pressure of the helium gas released to vacuum is measured by a mass spectrometer. The records of the partial pressure of He as a function of temperature are called Thermal Desorption Curves of helium. NRA uses nuclear reactions between a diagnostic beam accelerated to high energy and a target isotope in the sample. We think this is the most suitable analysis method for He-3 identification, where the nuclear reaction between deuterons accelerated to high energies and He-3 is used (d + $^3$He → $^4$He + p + 18.354 MeV) [13]. This produces alpha rays and protons. The protons in particular reach energies of above 14 MeV and can be easily ejected from the sample, making it possible to measure them with some conventional radiation detectors (CR-39 plates or SSBD).

TDS has been widely applied for various kind of metals to analyze the thermal desorption behavior of helium after helium ion implantation. Retention of He in Ni was examined using 20 keV helium ion whose range is about 0.1 μm, almost 100% captured at room temperature at low fluences. It was confirmed that the most of helium are retained in Ni up to 900°C at low fluence [14]. Yamauchi et al. studied also Cu, where almost 100% of the injected helium was kept up to 850°C at low fluences [15,16]. If the fluence is below $10^{17}$ ions/cm$^2$, Ni and Cu retain helium stably inside even when the temperature is raised to about 500 °C, which higher than the heat treatment temperature of the Cu-Ni/zirconia samples.

NRA analysis has been carried out for $ZrO_2$: about half of the He-3 is lost out of the sample at a temperature rise to 830°C in the analysis of Trocellier et al. [17]; about 70% is lost at a temperature rise to 927°C in the analysis of Costantini et al. [18]. $ZrO_2$ is a kind of ceramic. Therefore, the effect of microcracks in particular on helium retention properties would need to be considered. However, if the crystalline structure of $ZrO_2$ is poorly defective, it would be expected that almost 100% of He-3 would remain in the sample at a temperature rise of about 500°C.

Namely, for the post-heated the Cu-Ni/zirconia samples, at least in the metallic phase of Cu-Ni zone, it can be expected that almost all of the He-3 produced will remain in the sample. The Cu-Ni/zirconia sample we used in this study was 0.1 g for both TDS and NRA.

### 3.2. TDS

In the present Thermal Desorption Spectrometry (TDS) experiments, samples were subjected to a controlled temperature increase from ambient to 1200°C at a constant rate of 10°C/min. Partial pressures of various masses were measured using a quadrupole mass spectrometer (PrismaPro QMG250M2, Pfeiffer Vacuum, Germany). An infrared induction heater (THERMO RIKO CO., LTD., Japan) was employed to minimize unintended temperature increases beyond the sample and its holder (GVL298, THERMO RIKO, Japan).

He-3 signals manifest at mass-3; however, $H_3^+$ ions, inevitably produced in the mass spectrometer's ion source due to residual hydrogen-containing gases, also contribute to this signal. Despite this overlap, the thermal release behaviors of hydrogen and helium into vacuum differ significantly. Moreover, $H_3^+$ production, resulting from the reaction of hydrogen atoms (H or $H^+$) with $H_2$, is expected to correlate with the partial pressures of H, $H_2$, and $H_2O$. Conversely, the He-3 signal should not exhibit such correlations with hydrogen-containing species. The mass-3 signal may also include contributions from $DH^+$ (deuterium and protium molecular ion), which is anticipated to exhibit a temperature dependence similar to $H_2^+$, with release occurring in a temperature range distinct from helium. Tritium ($T^+$) could theoretically contribute to the mass-3 signal but can be disregarded unless intentionally introduced into the experimental system. While TDS allows for both qualitative and quantitative analysis in principle, this study focused solely on qualitative assessment.

## 3.3. NRA

The Nuclear Reaction Analysis (NRA) experiments were conducted using the M15 beam line of the tandem electrostatic accelerator (5SDH-2, National Electrostatics Corp. USA) at The Tandem Accelerator Laboratory of Kobe University (TAcLKU), Graduate School of Maritime Sciences, Japan. Deuteron beams with an energy of 1.4 MeV and a current of 5 nA were employed, with each sample subjected to 5 hours of irradiation. Proton detection was achieved using CR-39 detectors [19], a highly sensitive solid-state nuclear track detector (TechnoTrak, Chiyoda Technol Co., Japan) [20].

The powdered Cu-Ni/zirconia sample was contained within a 9×9 mm$^2$ aperture in a fluorescent plate for beam monitoring, secured on both sides by Kapton thin films. Two CR-39 detectors were stacked, with the target-facing surface shielded by aluminum foil. This configuration allowed for selective detection of protons exceeding 10 MeV energy, as only these could penetrate the first 0.9 mm thick CR-39 sheet and reach the second detector. The Monte Carlo code PHITS was utilized for design and optimization of the irradiation and analysis system. Quantitative analysis of He-3 in the samples was performed based on the number of proton-induced etch-pits observed [21].

Figure 1 shows the results of a PHITS simulation of the interaction of a 1.4 MeV deuterium beam with a Cu-Ni/zirconia target. The deuterium beam scattered by the target has a certain spread, but this can be avoided by placing the CR-39 detectors in the right positions. Figure 2 shows the behavior of protons produced as a result of the reaction between He-3 and deuterium in the target, also from a PHITS simulation. An arrangement has been chosen where the protons produced are able to vertically incident on the CR-39 detector. The calculation assumes that He-3 is present in the sample in equal amounts to Cu and emphasize the amounts of protons produced. The energy evaluation of protons and other particles using the CR-39 detector is detailed in papers by Kanasaki et al [22, 23]. Etching was carried out using a 6 M KOH solution kept at 70°C for 5 h. The etch pits were observed using an optical microscope (VHX-5000, KEYENCE CORPORATION, Japan) at Radioisotopes Division, Research Facility Center for Science and Technology, Kobe University.

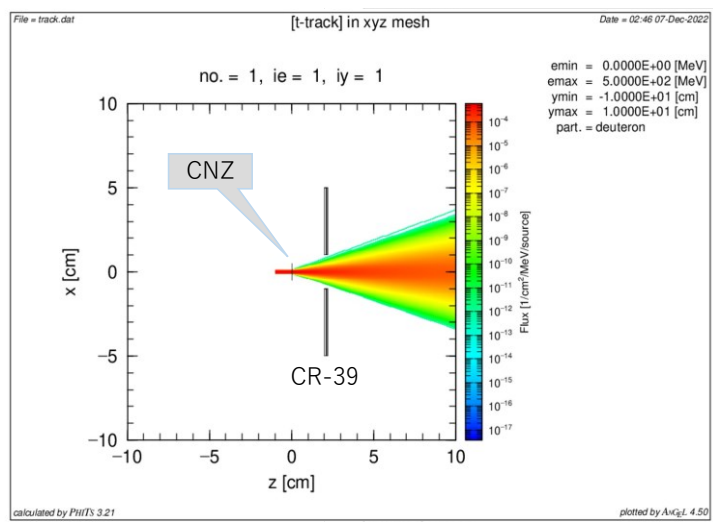

Fig. 1 Scattering behavior of 1.4 MeV deuterons by Cu-Ni/zirconia target (shown as CNZ in figure) from PHITS simulations.

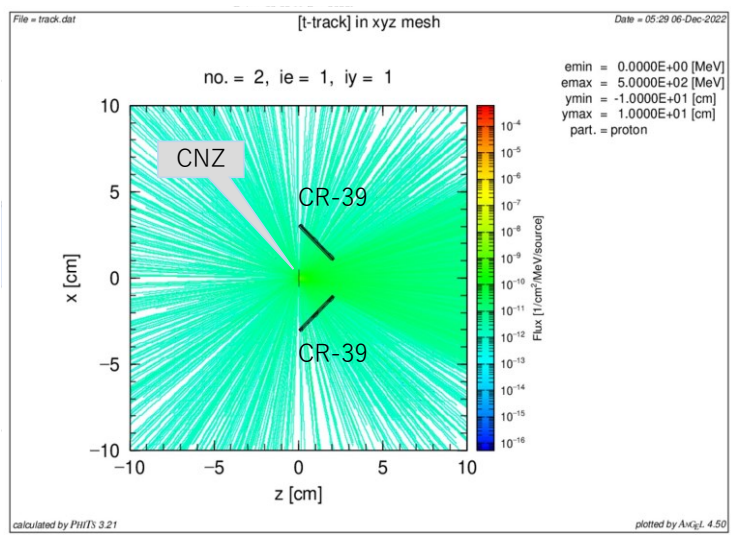

Fig. 2 PHITS simulation of proton production behavior from Cu-Ni/zirconia target (shown as CNZ in figure) irradiated with 1.4 MeV deuterons.

In the following sections, the results of the NRA, which were successfully analyzed quantitatively, are described first, followed by the TDS results. Three samples were prepared for the NRA and one sample was used for the TDS. The list is shown earlier in Table 1. The emphasis in selecting these samples was the high amount of He-3 expected to be present in them. As a result, samples with higher integrated calorific values were selected. Therefore, the exothermic experimental histories of the samples are not simple.

In addition to the initial exothermic experiment following the Cu-Ni/ZrO$_2$ preparation, the samples experienced two or three times exothermic experiments. The integrated heating values of the samples used in the NRA analysis differ from each other by a factor of more than 1.5. The initial composition ratios of samples #1 and #2 are different from those of sample #3, and the initial composition ratio of sample #4 for TDS is the same as that of sample #3. However, the ratios of total Cu and Ni to Zr are not significantly different, so the effect of the initial ratios on the analysis results might be not so significant. Note that each exothermic experiment used a sample of about 40 g, from which 0.1 g is collected for NRA or TDA. Each sample for analysis was taken from the area near the center of the exothermic samples, which were in close proximity to the heater. At least an error of 10% is expected in the calculations of calorific value, including weighing of the samples. Even if there were problems with the accuracy of the calculated values, the relative relationship between the magnitudes of the evaluated values would not be in error.

## 4. Results

### 4.1. Results from NRA

While surface-barrier semiconductor detectors (SSBD) are conventionally employed for proton measurement in Nuclear Reaction Analysis (NRA) [13], this study utilized CR-39 solid-state nuclear track detectors. The primary advantage of this approach lies in the creation of latent tracks—proton trajectories manifested as minute corrosion holes or etch pits—allowing for reliable detection and identification of individual protons. For near-vertical incidence of protons with energies below 10 MeV, CR-39 detectors exhibit a detection efficiency approaching 100% [20].

In the present experimental configuration, two CR-39 detectors were stacked. Protons decelerate while traversing the first CR-39 layer, resulting in only those with initial energies of approximately 10 MeV or greater reaching the second detector. The geometry and growth behavior of the etch pits, revealed through subsequent etching processes, provide unambiguous confirmation of their proton origin.

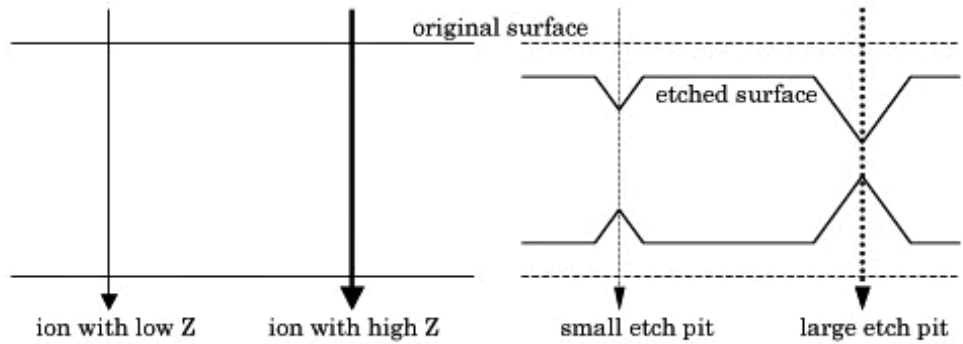

Fig. 3 Schematic diagram for the operation of an etched nuclear track detector. This is an example using relativistic high-energy particles, where Z is the nuclear charge.

Figure 3 schematically illustrates the operating principle of etched track detectors, depicting the case of two ion types penetrating the detector. In this scenario, the track etch rate ($V_t$) along the latent ion tracks exceeds the bulk etch rate ($V_b$). The geometry and dimensions of the resultant etch pits vary according to the charge and energy of the incident ions. For protons with energies surpassing the Bragg peak, an inverse relationship exists between energy and latent track damage density. Consequently, higher-energy protons induce lower track etching velocities, resulting in smaller etch pits. It is important to note that the etch pit geometry deviates from a perfect cone due to the depth-dependent variation in track etching rate. In the present study's configuration, protons with energies exceeding 10 MeV are expected to traverse the first CR-39 sheet before being arrested within the second, allowing for selective detection of high-energy protons.

Figure 4 shows an optical micrograph of the CR-39 surface after etching, representing the results of the NRA analysis for sample #3. Several etch-pits of different sizes are observed, all of which can be judged to be proton etch-pits from the size. The etch pits shown here are almost perpendicular to the detector surface and are unlikely to be neutron related. The possibility that they are naturally occurring alpha rays is ruled out both by the size of the etch pits and the fact that they are all almost vertically incident. Etch pits similar to those shown in this photograph were also observed for sample #1 and sample #2 with different number density. It is certain that the corresponding amount of He-3 exists in these samples.

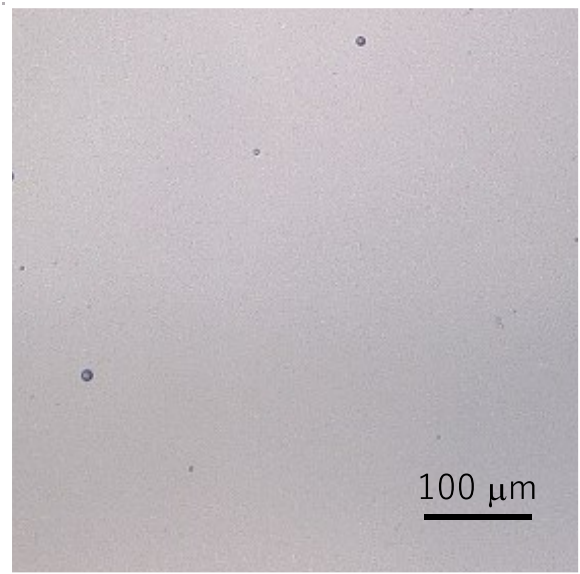

Fig. 4 Optical micrograph of the etch pit on the CR-39 track detector for the sample #3 listed in Table 1. The total calorific value was $5.5 \times 10^5$ J/0.1 g. The etch pit density on CR-39 was $1,113 \pm 23$ pits/cm$^2$.

Figure 5 shows the relationship between the integrated calorific value during the heat generation experiments in hydrogen gas and the number density of proton etch-pits recorded on the CR-39 for these three samples. The error in the abscissa heat value was derived from the heating value calculation in each step, while the error in the vertical etch pit density is each statistical error. It is clearly shown that the proton etch-pit density increases almost in proportion to the integrated calorific value. In addition, the best-fit straight line for the three experimental points passes approximately through the origin. This indicates that there is little influence other than the He-3 to produce proton etch-pits in the present experimental system.

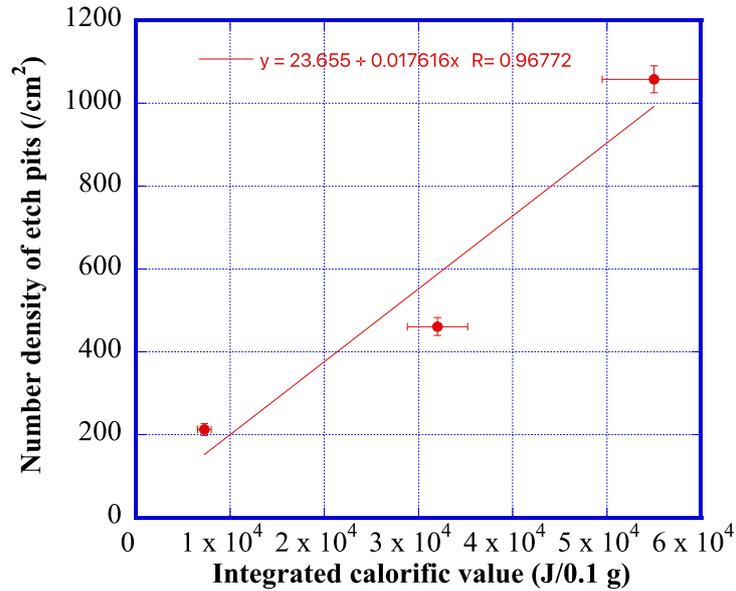

Fig. 5 Correlation between the Integrated calorific value of the Cu-Ni/zirconia samples and the number density of etch pits recorded on the CR-39 detector.

Based on the ratio of the number of protons expected to reach the CR-39 detector surface (10 - 16 MeV) to the measured number density of proton etch-pits in it, it was estimated that approximately $5.0\times10^{15}$ of He-3 was generated in the sample#3, using the Monte Carlo calculation code PHITS. The heat generation due to this would be estimated to be $6.2\times10^3$ J/0.1 g, when assuming that all of the unstable nuclei of Li*-4 produced undergo decays that emit He-3. This is one order of magnitude lower than the integrated calorific value shown in Table 1. Two factors may be influencing this discrepancy, although further investigation is required. One is the effect of the branch where deuterium is formed from the unstable Li*-4 nucleus, and the other may be the effect of the He-3 present in $ZrO_2$ being lost out of the sample during experiments and calcinations, where the mobility of He-3 might become high due to micro-crack introduction and other defects. Another possibility is contribution of heat generation by 4H/TSC induced fission processes to metal (Ni, Cu, Zr) nuclei [4], which can be further studied by foreign elements detection by conventional methods as TOF-SIMS, PIXE, etc.. This He-3 amount of $5\times10^{15}$ is at or slightly above the detection limit of normal mass spectrometry.

### 4.2. Results from TDS

Thermal desorption curves of the molecular species detected as a signal of mass-3 from the samples are shown in Fig. 6. The abscissa is the temperature measured by a thermocouple in contact with the top surface of the samples or sample dishes, while the output on the vertical axis is the combined signal of both He-3 and $H_3^+$. The possibility of $T^+$ (tritium ions) can be ruled out on good ground in this system. $DH^+$ should behave as like as $H_2^+$ ions under increasing temperature, and the total amount should be limited at the same level of He-3. At first, the focus is on desorption in the temperature range above 900°C. The indicating signal above 900 °C is also elevated for the sample dish plotted in blue and for the Cu-Ni/zirconia samples before the exothermic experiments in hydrogen gas plotted in red. It is apparent that the signals from Cu-Ni/zirconia samples after the exothermic experiments in hydrogen gas have higher values. The results of thermal desorption analyses of ion-implanted helium on Cu and Ni suggest that He-3 is released in this temperature range [14-16].

The main release of hydrogen gas $H_2$ from the Cu-Ni/zirconia samples (not shown here) is almost completely subdued by 800 °C, but becomes more pronounced again at temperatures above 900 °C. On the curve after the heating experiments in hydrogen gas, plotted in green, there is a high emission peak from 20°C to 100°C and several linked emission peaks from 100°C to 600°C. These were found to correlate well with the signal of molecular hydrogen $H_2$ (m/e = 2). This indicates that $H_3^+$ is produced dominantly by the reaction of $H_2$ with $H^+$ produced by the decomposition of $H_2$ in the ion source of the mass spectrometer, as discussed in the preliminary considerations. However, the signals of emissions in the temperature range above 900 °C showed poor correlation with $H_2$, making it difficult to attribute all of them to $H_3^+$. It is also unlikely that does $HD^+$ reach an amount that exceeds $H_3^+$ only in this temperature range.

As shown in Fig. 7, the signal of mass 3 from the emission above 800°C from the sample dish only and from the sample before the exothermic experiments in hydrogen gas showed a good correlation with the water signal (blue and red plots). In contrast to these, the signal from the sample#4 after the exothermic experiments in hydrogen gas increased independently of the increase in the water signal as the temperature increased, and even showed an inverse correlation when the temperature was increased further above 1100°C. This suggests the release of molecular species unrelated to hydrogen. Following a process of elimination, only the possibility of He-3 remains.

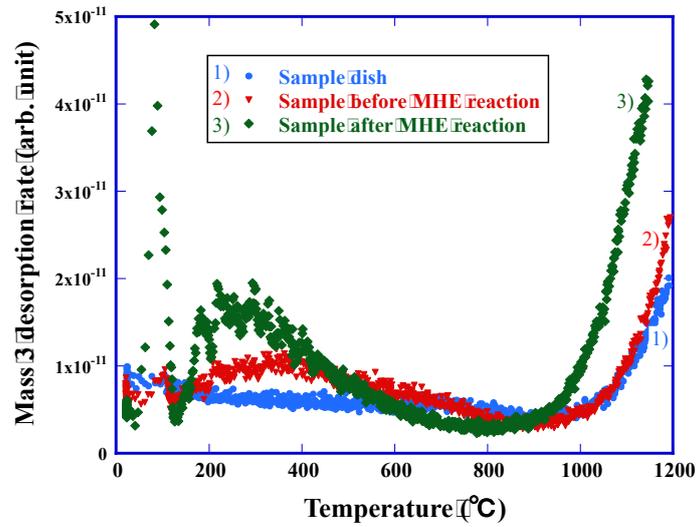

Fig. 6 Thermal desorption curves of He-3 and $H_3^+$ from 1) sample dish in bule, 2) sample before the experiment in red, and 3) sample#4, after the experiment in $H_2$ gas in green.

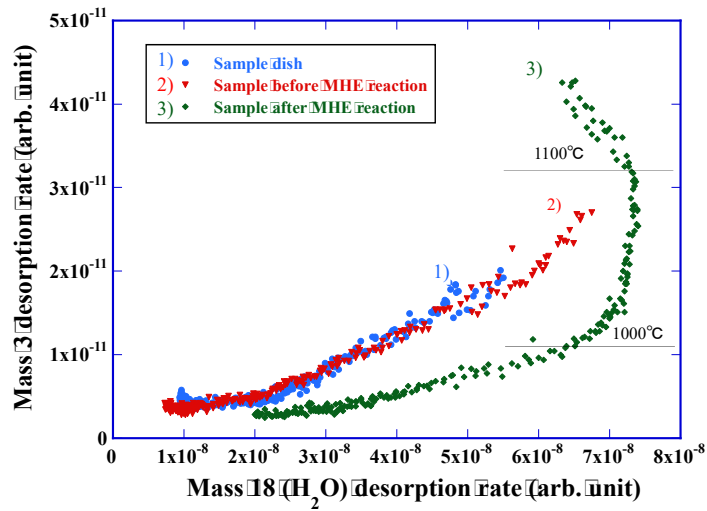

Fig. 7 Correlation between the mass 18 signal ($H_2O$) and the mass 3 signal (He-3 and $H_3^+$) at the temperature range above 800°C, from 1) sample dish in bule, 2) sample before the experiment in red, and 3) sample#4 in green.

## 5. Conclusion

This study demonstrates the detection of helium-3 in Cu-Ni/zirconia samples following exothermic experiments in hydrogen gas, utilizing two distinct analytical methods:

Nuclear Reaction Analysis (NRA) with deuteron beams and Thermal Desorption Spectrometry (TDS). The presence of helium-3, an extremely rare isotope in the natural environment, strongly suggests the occurrence of nuclear reactions within the Cu-Ni/zirconia samples. Given that the samples were exposed to normal hydrogen gas at elevated temperatures, the formation of helium-3 is likely attributable to nuclear fusion of hydrogen atoms.

These findings lend support to the 4H/TSC (4 Hydrogen/Tetrahedral Symmetric Condensate) model proposed by Emeritus Professor Akito Takahashi of Osaka University, which predicts helium-3 as a primary reaction product. While there exists a discrepancy of approximately one order of magnitude between the experimentally observed excess heat generation and estimated helium-3 production, these results remain consistent with the 4H/TSC theory when considering potential alternative decay processes of unstable Li*-4 that do not yield helium-3, as well as the possible degradation of helium-3 retention properties in $ZrO_2$ during heating experiments and contribution of nuclear heat source by 4H/TSC induced metal nuclei fission processes.

Future research directions should focus on elucidating the quantitative relationship between calorific value and helium-3 production. This can be achieved by expanding the number of cases analyzed via NRA and establishing quantitative analysis methods for TDS, particularly for samples exhibiting higher calorific values than those presently studied.

This investigation represents a significant milestone in the field of cold fusion research, which originated with the seminal work of Fleischmann and Pons in 1989. The results presented herein provide compelling evidence that cold fusion is not a theoretical construct but an observable phenomenon, offering potential as a carbon-free and radioactivity-free energy source for humanity.

Further advancements in fusion reaction efficiency may be realized through refinements to Cu-Ni/zirconia materials. Immediate research priorities include defining optimal conditions for Cu-Ni/zirconia sample preparation using existing methods, followed by exploration of alternative fabrication techniques. Additionally, the development of novel hydrogen fusion materials based on innovative concepts warrants investigation.

This research opens avenues for the development of a new nuclear technology paradigm, distinct from conventional approaches rooted in atomic and hydrogen bomb development. Such advancements hold the promise of addressing global energy challenges while potentially mitigating societal tensions associated with traditional nuclear technologies.

**Acknowledgements**

The authors would like to thank to all staff of The Tandem Accelerator Laboratory of Kobe University, TAcLKU, for their warm supports during the experiments. The present work is supported by The Thermal & Electronic Energy Technology (TEET) Foundation, Kobe University Innovation Found Program, and JST START University Ecosystem Promotion Type (University Promotion Type), Grant Number JPMJST2051, Japan. The authors thank Professor C. Gomez for fruitful discussions, including perspectives from the philosophy of science, and for his help in improving our English language style.